\documentclass[11pt]{article}
\usepackage{amsmath,amsfonts,latexsym,graphicx,amssymb}
\begin{document}

\def\theequation{\thesection.\arabic{equation}}

\def\llra{\relbar\joinrel\longrightarrow}              
\def\mapright#1{\smash{\mathop{\llra}\limits_{#1}}}    
\def\mapup#1{\smash{\mathop{\llra}\limits^{#1}}}     
\def\mapupdown#1#2{\smash{\mathop{\llra}\limits^{#1}_{#2}}} 

\title{\bf The Rigged Hilbert Space of the Free Hamiltonian}

\author{Rafael de la Madrid \\ [1ex]
\small{\it Institute for Scientific Interchange (ISI), 
Villa Gualino, Viale Settimio Severo 65} \\ [-0.5ex]
\small{\it I-10133, Torino, Italy} \\ [-0.5ex]
\small{\it E-mail: \texttt{rafa@isiosf.isi.it}}}

\date{\small{(October 23, 2003)}}

\maketitle

\begin{abstract}
\noindent We explicitly construct the Rigged Hilbert Space (RHS) of the free 
Hamiltonian $H_0$. The construction of the RHS of $H_0$ provides yet another 
opportunity to see that when continuous spectrum is present, the solutions of 
the Schr\"odinger equation lie in a RHS rather than just in a Hilbert space.
\end{abstract}

\vskip0.3cm



\newpage

\section{Introduction}
\setcounter{equation}{0}
\label{sec:introduction}

There is a growing realization that the Rigged Hilbert Space (RHS) provides 
the methods needed to handle Dirac's bra-ket formalism and continuous 
spectrum. Moreover, there is an increasing number of Quantum Mechanics 
textbooks that include the RHS as part of their 
contents~\cite{ATKINSON}--\cite{GALINDO} (this list is 
non-exhaustive). However, there 
is still a lack of simple examples for which the RHS is explicitly constructed 
(one exception is Ref.~\cite{B70}). Especially important is to construct the 
RHS generated by the Schr\"odinger equation, because claiming that Quantum 
Mechanics needs the RHS is tantamount 
to claiming that the solutions of the Schr\"odinger equation lie in a RHS (when
continuous spectrum is present). The task of constructing the RHS generated by
the Schr\"odinger equation was undertaken in Ref.~\cite{DIS}. The method
proposed in \cite{DIS} has been applied to two simple 
potentials~\cite{JPA,FP02}. In this paper, we shall apply this method 
to the simplest example possible: the free Hamiltonian. 

We note that the results of this paper follow immediately from those in 
Refs.~\cite{JPA,FP02}, by making the value of the potential zero. Nevertheless,
we think that the example of the free Hamiltonian provides a very 
transparent way to understand the essentials of the method of \cite{DIS}, 
because the calculations are reduced to the minimum. 

The time-independent Schr\"odinger equation for the free Hamiltonian $H_0$
reads, in the position representation, as
\begin{equation}
	\label{delta}
	\langle \vec{x}|H_0|E\rangle =
	\frac{-\hbar^2}{2m} \nabla^2
	\langle \vec{x}|E\rangle =E\langle \vec{x}|E\rangle \, ,
\end{equation}
where $\nabla^2$ is the three-dimensional Laplacian. In spherical coordinates 
$\vec{x}\equiv (r,\theta, \phi)$, Eq.~(\ref{delta}) has the following form:
\begin{equation}
	\langle r,\theta, \phi|H_0|E,l,m \rangle =
	\biggl( \frac{-\hbar^2}{2m}\frac{1}{r}\frac{\partial^2}{\partial
r^2}r
	+\frac{\hbar^2l(l+1)}{2mr^2} \biggr)
 \langle r,\theta, \phi|E,l,m\rangle 
 = E\langle r,\theta, \phi|E,l,m\rangle\,.
	\label{sphSe}
\end{equation}
By separating the radial and angular dependences, 
\begin{equation}
	\langle r,\theta, \phi|E,l,m\rangle \equiv
	\langle r|E\rangle_l \, \langle \theta, \phi |l,m\rangle \equiv
	\frac{1}{r}\chi_l(r;E)Y_{l,m}(\theta, \phi),
\end{equation}
where $Y_{l,m}(\theta, \phi)$ are the spherical harmonics, we obtain for 
the radial part
\begin{equation}
\biggl(\frac{-\hbar^2}{2m}\frac{d^2}{dr^2}+\frac{\hbar^2l(l+1)}{2mr^2}
	\biggr) \chi_l(r;E)=E\chi_l(r;E)\,.
	\label{baba}
\end{equation}
In this paper, we shall restrict ourselves to the case of zero orbital 
angular momentum (the higher-order case can be treated analogously). We then 
write $\chi_{l=0}(r;E)\equiv \chi (r;E)$ and obtain
\begin{equation}
	-\frac{\hbar^2}{2m} \frac{d^2}{dr^2}\chi(r;E)=
         E\chi (r;E)\,.
	\label{rSe0}
\end{equation}
We shall write this equation as
\begin{equation}
	h_0\chi(r;E)= E\chi (r;E) \, ,
	\label{rSe0s}
\end{equation}
where 
\begin{equation}
      h_0\equiv -\frac{\hbar ^2}{2m}\frac{d^2}{dr^2} 
      \label{0doh}
\end{equation}
is the formal differential operator corresponding to the free Hamiltonian 
(for $l=0$). Our goal is to solve Eq.~(\ref{rSe0s}) and show that its 
solutions lie in a RHS rather than just in a Hilbert space. 

The basic tool necessary to solve Eq.~(\ref{rSe0s}) is the Sturm-Liouville 
theory~\cite{SL}. This theory provides the Hilbert space 
methods. As shown in many publications (cf.~\cite{DIS} and references
therein), the Hilbert space methods do not provide us with all the tools 
needed in Quantum Mechanics when continuous spectrum is present. In particular,
the Hilbert space cannot incorporate Dirac's bra-ket formalism. Therefore,
an extension of the Hilbert space is needed. The extension that seems to be
most suitable is the RHS (cf.~\cite{DIS} and references therein). In 
particular, the RHS incorporates Dirac's bra-ket formalism.

The structure of the paper is as follows. In Section~\ref{sec:iseaexte}, we
construct the domain and the self-adjoint extension of the differential
operator (\ref{0doh}). In Section~\ref{sec:resoangf}, we obtain the free 
Green function, whose expression is used in Section~\ref{sec:sphspo} to
calculate the spectrum of $H_0$. Section~\ref{sec:diaaeiexp} is devoted to 
the eigenfunction expansion and the direct integral decomposition of the
Hilbert space. In Section~\ref{0sec:tneodRHS}, we construct the RHS of $H_0$.
The Dirac basis vector expansion for $H_0$ is obtained in 
Section~\ref{sec:0DBVE}, and the energy representation of the RHS of $H_0$
is constructed in Section~\ref{sec:0ESrepr}. Finally, the results of the
paper are summarized in the diagram of Eq.~(\ref{0diagramsavp}).

\section{Self-Adjoint Extension}
\setcounter{equation}{0}
\label{sec:iseaexte}

The first step is to define a linear operator on
a Hilbert space corresponding to the formal differential operator 
(\ref{0doh}). In the radial position representation, the Hilbert space that
belongs to the RHS of the free Hamiltonian is 
realized by the space $L^2([0,\infty ), dr)$ of square integrable functions 
$f(r)$ defined on the interval $[0,\infty )$ (see the diagram 
(\ref{0diagramsavp}) below). 

The domain ${\cal D}(H_0)$ of the free Hamiltonian must be a proper 
dense linear subspace of $L^2([0,\infty ), dr)$. The action of $h_0$ must be 
well defined on ${\cal D}(H_0)$, and this action must remain 
in $L^2([0,\infty ), dr)$. We need also a boundary condition that assures 
the self-adjointness of the Hamiltonian. The boundary conditions that 
select the possible self-adjoint extensions of $h_0$ are given by (see
\cite{DS}, page 1306)
\begin{equation}
       f(0)+\alpha  \, f'(0)=0 \, ,
       \quad -\infty < \alpha \leq \infty \, .
\end{equation}
Among all these boundary conditions, we 
choose $f(0)=0$. Therefore, the requirements that are to be fulfilled by the
elements of the domain of $H_0$ are 
\begin{subequations}
      \label{bcthdpspr}
\begin{eqnarray}
      &&f(r) \in L^2([0,\infty ), dr)  \, , \label{HSc} \\
      &&h_0f(r) \in L^2([0,\infty ), dr) \, , \label{reduce} \\
       &&f(r) \in AC^2 [0,\infty ) \, ,  \label{ACc} \\
       &&f(0)=0 \, , \label{sac}  
\end{eqnarray}
\end{subequations}
where $AC^2[0,\infty)$ denotes the space of functions whose derivative 
is absolutely continuous (for details on absolutely continuous functions,
consult Refs.~\cite{DS,DIS}). The requirements in 
Eq.~(\ref{bcthdpspr}) yield the domain of $H_0$:
\begin{equation}
       {\cal D}(H_0) =\{ f(r)\, | \ f(r), h_0f(r)\in L^2([0,\infty ), dr), \,
                   f(r) \in AC^2[0,\infty ), \, f(0)=0 \} \, .
     \label{0domain}
\end{equation}
On ${\cal D}(H_0)$ the formal differential operator $h_0$ is self-adjoint. In 
choosing (\ref{0domain}) as the domain of our formal differential
operator $h_0$, we define a linear operator $H_0$ by
\begin{equation}
      H_0f(r) :=h_0f(r)=-\frac{\hbar ^2}{2m}\frac{d^2}{dr^2}f(r)
       \, , \quad f(r) \in {\cal D}(H_0) \, .
      \label{0operator}
\end{equation}

\section{Resolvent and Green Function}
\setcounter{equation}{0}
\label{sec:resoangf}

The expression of the free Green function $G_0(r,s;E)$ is given in terms 
of eigenfunctions of the differential operator $h_0$ subject to certain
boundary conditions (cf.~Theorem~1 in Appendix~\ref{sec:A1}). We shall
divide the complex energy plane in three regions, and calculate
$G_0(r,s;E)$ for each region separately. In all our
calculations, we shall use the following branch of the square root function:
\begin{equation}
      \sqrt{\cdot}:\{ E\in {\mathbb C} \, | \  -\pi <{\rm arg}(E)\leq \pi \} 
   \longmapsto \{E\in {\mathbb C} \, | \  -\pi/2 <{\rm arg}(E)\leq \pi/2 \} 
      \, .
   \label{branch} 
\end{equation}

\def\thesubsection{\thesection.\arabic{subsection}}
\subsection{Region $\mbox{Re}(E)<0$, $\mbox{Im}(E)\neq 0$}
\label{sec:relolo}

For $\mbox{Re}(E)<0$, $\mbox{Im}(E)\neq 0$, the free Green function (see 
Theorem~1 in Appendix~\ref{sec:A1}) is given by
\begin{equation}
       G_0(r,s;E)=\left\{ \begin{array}{ll}
      -\frac{2m/\hbar ^2}{\sqrt{-2m/\hbar ^2 \, E}} \,
      \frac{\widetilde{\chi}(r;E) \, \widetilde{f}(s;E)}{2}
      \quad  &r<s \\ [2ex]
     -\frac{2m/\hbar ^2}{\sqrt{-2m/\hbar ^2 \, E}}\, 
   \frac{\widetilde{\chi}(s;E)\,  \widetilde{f}(r;E)}{2}
     \quad &r>s 
        \end{array} 
      \right. 
      \quad \mbox{Re}(E)<0 \, , \ \mbox{Im}(E)\neq 0 \, .
	\label{0-}
\end{equation}
The eigenfunction $\widetilde{\chi}(r;E)$ satisfies the Schr\"odinger 
equation, Eq.~(\ref{rSe0s}), and the boundary conditions 
\begin{subequations}
\begin{eqnarray}
       && \widetilde{\chi} (0;E)=0 \, , \label{bca01} \\
       && \widetilde{\chi}(r;E)
           {\rm \ is \ square \ integrable \ at \ } 0 \, ,
       \label{sbca03} 
\end{eqnarray}
      \label{eigsbo0co}
\end{subequations}
which yield
\begin{equation}
      \widetilde{\chi}(r;E) =e^{\sqrt{-\frac{2m}{\hbar ^2}E} \, r}-
                e^{-\sqrt{-\frac{2m}{\hbar ^2}E} \, r} \, , 
                \quad 0<r<\infty \, .
       \label{0tildechifunction}
\end{equation}
The eigenfunction $\widetilde{f}(r;E)$ satisfies the Schr\"odinger equation, 
Eq.~(\ref{rSe0s}), and the boundary condition
\begin{equation}
       \widetilde{f}(r;E) \ 
         {\rm is \ square \ integrable \ at \ } \infty \, ,
         \label{thetcoabejej}
\end{equation}
which yield
\begin{equation}
      \widetilde{f}(r;E)=e^{-\sqrt{- \frac{2m}{\hbar ^2}E}\, r} \, ,
                 \quad  0<r<\infty \, . 
      \label{0tildethetfunc}
\end{equation}
The Wronskian of $\widetilde{\chi}$ and $\widetilde{f}$ can be easily
calculated:
\begin{equation}
       W(\widetilde{\chi},\widetilde{f})=-2 \sqrt{- \frac{2m}{\hbar ^2}E} \, .
\end{equation}

\def\thesubsection{\thesection.\arabic{subsection}}
\subsection{Region $\mbox{Re}(E)>0$, $\mbox{Im}(E)>0$}
\label{sec:regogo}

When $\mbox{Re}(E)>0$, $\mbox{Im}(E)>0$, the expression of the free Green 
function is
\begin{equation}
       G_0(r,s;E)=\left\{ \begin{array}{ll}
       -\frac{2m/\hbar ^2}{\sqrt{2m/\hbar ^2 \, E}} \,
         \chi (r;E) \, f^+ (s;E) \quad
         &r<s  \\ [2ex]
       -\frac{2m/\hbar ^2}{\sqrt{2m/\hbar ^2 \, E}} \,
         \chi (s;E) \, f^+ (r;E) \quad
               &r>s 
         \end{array} 
                 \right. 
       \quad \mbox{Re}(E)>0, \ \mbox{Im}(E)>0 \, . 
	\label{0++}
\end{equation}
The eigenfunction $\chi (r;E)$ satisfies Eq.~(\ref{rSe0s}) and the boundary 
conditions (\ref{eigsbo0co}), which yield
\begin{equation}
      \chi (r;E)=\sin ( \sqrt{\frac{2m}{\hbar ^2}E}\, r ) \, , 
      \quad 0<r<\infty \, .
      \label{0chi}
\end{equation}
The eigenfunction $f^+(r;E)$ satisfies Eq.~(\ref{rSe0s}) subject to the 
boundary condition (\ref{thetcoabejej}), which yield
\begin{equation}
      f^{+}(r;E)=e^{i\sqrt{\frac{2m}{\hbar ^2}E}\, r} \, ,
                 \quad  0<r<\infty \, . 
      \label{0theta+fun}
\end{equation}
The expression of the Wronskian of $\chi$ and $f^+$ follows immediately from
Eqs.~(\ref{0chi}) and (\ref{0theta+fun}):
\begin{equation}
       W(\chi ,f^+)=- \sqrt{\frac{2m}{\hbar ^2}E} \, .
\end{equation}

\def\thesubsection{\thesection.\arabic{subsection}}
\subsection{Region $\mbox{Re}(E)>0$, $\mbox{Im}(E)< 0$}
\label{sec:regolo}

In the region $\mbox{Re}(E)>0$, $\mbox{Im}(E)<0$, the free Green function reads
\begin{equation}
       G_0(r,s;E)=\left\{ \begin{array}{ll}
       -\frac{2m/\hbar ^2}{\sqrt{2m/\hbar ^2 \, E}} \,
         \chi (r;E) \, f^- (s;E) \quad
         &r<s  \\ [2ex]
       -\frac{2m/\hbar ^2}{\sqrt{2m/\hbar ^2 \, E}} \,
         \chi (s;E) \, f^- (r;E) \quad 
               &r>s 
         \end{array} 
                 \right. 
       \quad \mbox{Re}(E)>0, \ \mbox{Im}(E)<0 \, . 
	\label{0+-}
\end{equation}
The eigenfunction $\chi (r;E)$ is given by (\ref{0chi}), although now
$E$ belongs to the fourth quadrant of the energy plane. The 
eigenfunction $f^-(r;E)$ satisfies Eq.~(\ref{rSe0s})
and the boundary condition (\ref{thetcoabejej}), which yield
\begin{equation}
      f^-(r;E)=e^{-i\sqrt{\frac{2m}{\hbar ^2}E}\, r} \, ,
                 \quad  0<r<\infty \, . 
      \label{0thetafun-}
\end{equation}
The Wronskian of $\chi$ and $f^-$ is given by
\begin{equation}
       W(\chi ,f^-)=- \sqrt{\frac{2m}{\hbar ^2}E} \, .
\end{equation}

\section{Spectrum of $H_0$}
\setcounter{equation}{0}
\label{sec:sphspo}

In this section, we obtain the spectrum of $H_0$, which we shall
denote by ${\rm Sp}(H_0)$. We know
that $H_0$ is self-adjoint, and therefore its spectrum is real. In order to 
obtain the set of real numbers that belong to ${\rm Sp}(H_0)$, we apply
Theorem~4 of Appendix~\ref{sec:A1}. From Theorem~4 and from the previous 
section, it is clear that we should study the positive and the negative
real lines separately. As expected, we shall obtain that 
${\rm Sp}(H_0)=[0,\infty )$.

\def\thesubsection{\thesection.\arabic{subsection}}
\subsection{Subset $\Lambda =(-\infty ,0)$}
\label{sec:sulsos}

We first take $\Lambda$ from Theorem~4 of Appendix~\ref{sec:A1} to be 
$(-\infty ,0)$. We choose a basis for the space of solutions of the 
equation $h_0\sigma =E\sigma$ as
\begin{subequations}
\begin{eqnarray}
      &&\widetilde{\sigma}_1(r;E)=e^{\sqrt{-\frac{2m}{\hbar ^2}E} \, r}\, ,
             \qquad  \label{0tildesigma1} \\  [2ex]
      &&\widetilde{\sigma}_2(r;E)=\widetilde{f}(r;E) \, .
\end{eqnarray}
\end{subequations}

Obviously,
\begin{equation}
      \widetilde{\chi}(r;E)=\widetilde{\sigma}_1(r;E)-
     \widetilde{\sigma}_2(r;E) \, ,
\end{equation}
which along with Eq.~(\ref{0-}) leads to
\begin{equation}
      G_0(r,s;E)= 
      -\frac{2m/\hbar ^2}{\sqrt{-2m/\hbar ^2 \, E}} \,
       \frac{1}{2}\, \left[ \widetilde{\sigma}_1(r;E)- 
       \widetilde{\sigma}_2(r;E)\right]\widetilde{\sigma}_2(s;E) \, , 
       \  r<s \, , \  \mbox{Re}(E)<0 \, , \mbox{Im}(E) \neq 0 \, . 
        \label{0dis-parosof}
\end{equation}
Because
\begin{equation}
      \overline{\widetilde{\sigma}_2(s;\overline{E})}=
      \widetilde{\sigma}_2(s;E) \, ,
\end{equation}
we can write Eq.~(\ref{0dis-parosof}) as
\begin{eqnarray}
       &&G_0(r,s;E)= 
      -\frac{2m/\hbar ^2}{\sqrt{-2m/\hbar ^2 \, E}} \,
       \frac{1}{2}\,  \left[ \widetilde{\sigma}_1(r;E)
       \overline{\widetilde{\sigma}_2(s;\overline{E})}
       -\widetilde{\sigma}_2(r;E)
       \overline{\widetilde{\sigma}_2(s;\overline{E})} \right] , \nonumber \\
       &&\qquad \hskip6cm
        r<s \, , \  \mbox{Re}(E)<0 \, , \mbox{Im}(E) \neq 0 \, .
        \label{0finag-retoco} 
\end{eqnarray}
On the other hand, by Theorem~4 in Appendix~\ref{sec:A1} we have
\begin{equation}
       G_0(r,s;E)=\sum_{i,j=1}^{2}
       \theta _{ij}^- (E)\widetilde{\sigma}_i(r;E)
        \overline{\widetilde{\sigma}_j(s;\overline{E})} \, ,
       \qquad r<s\, .
       \label{0gf--}
\end{equation}
By comparing Eqs.~(\ref{0finag-retoco}) and (\ref{0gf--}) we see that
\begin{equation}
      \theta _{ij}^-(E)= \left(  \begin{array}{cc}
        0 & -\frac{2m/\hbar ^2}{\sqrt{-2m/\hbar ^2 \, E}} 
       \frac{1}{2}  \\
        0 &
      \frac{2m/\hbar ^2}{\sqrt{-2m/\hbar ^2 \, E}} 
       \frac{1}{2}   
                                \end{array}
        \right) , \quad \mbox{Re}(E)<0 \, , \  \mbox{Im}(E) \neq 0 \, .
\end{equation}
The functions $\theta _{ij}^-(E)$ are analytic in a neighborhood of 
$\Lambda =(-\infty ,0)$. Therefore, the interval $(-\infty ,0)$ is in the 
resolvent set, ${\rm Re}(H_0)$, of the operator $H_0$.

\def\thesubsection{\thesection.\arabic{subsection}}
\subsection{Subset $\Lambda =(0, \infty )$}
\label{sec:sulsosge}

In this case, we choose the following basis for the space of solutions 
of $h_0\sigma =E\sigma$:  
\begin{subequations}
      \label{0basisin0infisi}
\begin{eqnarray}
      &&\sigma _1(r;E)=\chi (r;E)\, ,
      \label{0sigma1=sci} \\ [2ex]
      &&\sigma _2(r;E)=\cos (\sqrt{\frac{2m}{\hbar ^2}E} \, r) \, .
      \label{0sigam2cos}  
\end{eqnarray}
\end{subequations}

Eqs.~(\ref{0theta+fun}), (\ref{0thetafun-}) and (\ref{0basisin0infisi}) lead 
to  
\begin{equation}
      f^+(r;E)=i \sigma _1(r;E)+ \sigma _2(r;E) 
      \label{0Thet+inosigm}
\end{equation}
and to
\begin{equation}
      f^-(r;E)=-i\sigma _1(r;E)+\sigma _2(r;E) \, .
      \label{0Thet-inosigm}
\end{equation}
By substituting Eq.~(\ref{0Thet+inosigm}) into Eq.~(\ref{0++}) we get to
\begin{equation}
      G_0(r,s;E)=
       -\frac{2m/\hbar ^2}{\sqrt{2m/\hbar ^2 \, E}} \,
       \sigma _1(s;E)\left[ i\sigma _1 (r;E)+\sigma _2(r;E)\right] 
        , \ r>s  \, , \  \mbox{Re}(E)>0, \mbox{Im}(E)>0 \, .
        \label{0G++tofpuon}
\end{equation}
By substituting Eq.~(\ref{0Thet-inosigm}) into
Eq.~(\ref{0+-}) we get to
\begin{equation}
      G_0(r,s;E)=
      -\frac{2m/\hbar ^2}{\sqrt{2m/\hbar ^2 \, E}}   \,
      \sigma _1(s;E) \left[ -i\sigma _1(r;E)+\sigma _2(r;E) \right]      
      , \  r>s \, , \      \mbox{Re}(E)>0, \mbox{Im}(E)<0\, .
      \label{0G+-tofpuon}
\end{equation} 
Because
\begin{equation}
      \overline{\sigma _1(s;\overline{E})}=\sigma _1(s;E) \, ,
\end{equation}
Eq.~(\ref{0G++tofpuon}) leads to
\begin{eqnarray}
      &&G_0(r,s;E)=
      -\frac{2m/\hbar ^2}{\sqrt{2m/\hbar ^2 \, E}} \,
      \left[ i\sigma _1(r;E)\overline{\sigma _1(s;\overline{E})}   
      +\sigma _2(r;E)\overline{\sigma _1(s;\overline{E})}\right] , 
      \nonumber \\      
      &&\qquad \hskip6cm \mbox{Re}(E)>0, \mbox{Im}(E)>0 , \, r>s \, ,
       \label{0redaot++}
\end{eqnarray}
whereas Eq.~(\ref{0G+-tofpuon}) leads to 
\begin{eqnarray}
      &&G_0(r,s;E)=
      -\frac{2m/\hbar ^2}{\sqrt{2m/\hbar ^2 \, E}} \,
       \left[ -i\sigma _1(r;E)\overline{\sigma _1(s;\overline{E})}
        +\sigma _2(r;E)\overline{\sigma _1(s;\overline{E})}\right] , 
         \nonumber \\
       &&\qquad \hskip6cm \mbox{Re}(E)>0, \mbox{Im}(E)<0\, , \, r>s \, .
       \label{0redaot+-}
\end{eqnarray}
The expression of the resolvent in terms of the basis $\sigma _1,\sigma _2$
can be written as (see Theorem~4 in Appendix~\ref{sec:A1})
\begin{equation}
       G_0(r,s;E)=\sum_{i,j=1}^{2}
       \theta _{ij}^+ (E)\sigma _i(r;E)\overline{\sigma _j(s;\overline{E})}\, ,
       \qquad r>s \, .
        \label{0GF++}
\end{equation}
By comparing (\ref{0GF++}) to (\ref{0redaot++}) we get to
\begin{equation}
      \theta _{ij}^+(E)= \left(  \begin{array}{cc}
      -\frac{2m/\hbar ^2}{\sqrt{2m/\hbar ^2 \, E}} i
       & -\frac{2m/\hbar ^2}{\sqrt{2m/\hbar ^2 \, E}} \\
       0 & 0
                                \end{array}
        \right) 
       , \quad  \mbox{Re}(E)>0 \, , \  \mbox{Im}(E)>0 \, .
       \label{0theta++}
\end{equation}
By comparing (\ref{0GF++}) to (\ref{0redaot+-}) we get to
\begin{equation}
      \theta _{ij}^+(E)= \left(  \begin{array}{cc}
      \frac{2m/\hbar ^2}{\sqrt{2m/\hbar ^2 \, E}} i
       & -\frac{2m/\hbar ^2}{\sqrt{2m/\hbar ^2 \, E}} \\
       0 & 0
                                \end{array}
        \right) 
       , \quad  \mbox{Re}(E)>0 \, , \  \mbox{Im}(E)<0 \, .
      \label{0theta+-}
\end{equation}
From Eqs.~(\ref{0theta++}) and (\ref{0theta+-}) we can see that the measures 
$\varrho _{12}$, $\varrho _{21}$ and $\varrho _{22}$ in Theorem~4 of 
Appendix~\ref{sec:A1} are zero, and that the measure $\varrho _{11}$ is 
given by
\begin{eqnarray}
     \varrho _{11}((E_1,E_2))&=&\lim _{\delta \to 0} \lim _{\varepsilon \to 0+}
      \frac{1}{2\pi i} \int_{E_1+\delta}^{E_2-\delta}
      \left[ \theta _{11}^+ (E-i\varepsilon ) -\theta _{11}^+ (E+i\varepsilon )
      \right] dE \nonumber \\
      &=&\int_{E_1}^{E_2}   \frac{1}{\pi}\, 
      \frac{2m/\hbar ^2}{\sqrt{2m/\hbar ^2 \, E}}\, dE \, ,
\end{eqnarray}
which leads to
\begin{equation}
      \varrho (E)\equiv \varrho _{11}(E)=
      \frac{1}{\pi}\, 
      \frac{2m/\hbar ^2}{\sqrt{2m/\hbar ^2 \, E}}\, , 
      \quad E\in (0,\infty )  \, .
\end{equation} 
The function $\theta _{11}^+(E)$ has a branch cut along $(0,\infty)$, and 
therefore $(0,\infty )$ is included in ${\rm Sp}(H_0)$. Because 
${\rm Sp}(H_0)$ is a closed set, ${\rm Sp}(H_0)=[0,\infty )$.

\section{Diagonalization and Eigenfunction Expansion}
\setcounter{equation}{0}
\label{sec:diaaeiexp}

In the present section, we diagonalize $H_0$ and construct the 
expansion of the wave functions in terms of the eigenfunctions of the 
differential operator $h_0$. 

By Theorem~2 of Appendix~\ref{sec:A1}, there is a unitary map $\widetilde U_0$
defined by
\begin{eqnarray}
     \widetilde{U}_0:L^2([0,\infty ),dr) 
      &\longmapsto & L^2( (0,\infty ),\varrho (E)dE) \nonumber \\
       f(r)& \longmapsto & 
         \widetilde{f}(E)=\widetilde{U}_0f(E)=\int_0^{\infty}dr f(r) 
                             \overline{\chi (r;E)} \, ,
      \label{0rhoU}
\end{eqnarray}
that brings ${\cal D}(H_0)$ onto the space
\begin{equation}
      {\cal D}(\widetilde{H}_0)=\{ \widetilde{f}(E) \in 
                             L^2( (0,\infty ),\varrho (E)dE) \, | \  
                             \int_0^{\infty}dE \,
                              E^2|\widetilde{f}(E)|^2 \varrho (E)
                             <\infty \} \, .
       \label{0rhospace}
\end{equation}
The unitary operator $\widetilde U_0$ provides a $\varrho $-normalization
(cf.~Ref.~\cite{FP02}). In order to obtain a $\delta$-normalization, the 
measure 
$\varrho (E)$ must be absorbed in the definition of the eigenfunctions 
(cf.~Ref.~\cite{FP02}). This is why we define
\begin{equation}
      \sigma (r;E):=\sqrt{\varrho (E)} \, \chi (r;E) \, ,
      \label{0dnes}
\end{equation}
which is the $\delta$-normalized eigensolution of the differential operator 
$h_0$. If we define
\begin{equation}
       \widehat{f}(E):=\sqrt{\varrho (E)}{\widetilde f}(E) \, , \quad
       \widetilde{f}(E) \in L^2( (0,\infty ),\varrho (E)dE) \, ,
\end{equation}
and construct the unitary operator
\begin{eqnarray}
      \widehat{U}_0:L^2((0,\infty),\varrho (E)dE) &\longmapsto & 
               L^2((0,\infty),dE)
                                                           \nonumber \\
                        {\widetilde f} &\longmapsto & 
                        \widehat{f}(E)= \widehat{U}_0{\widetilde f}(E):=
                         \sqrt{\varrho (E)}{\widetilde f}(E) \, , 
\end{eqnarray}
then the operator that $\delta$-diagonalizes our Hamiltonian is
$U_0:={\widehat U}_0{\widetilde U}_0$,
\begin{eqnarray}
      U_0:L^2([0,\infty),dr) &\longmapsto & L^2((0,\infty),dE) \nonumber \\
                       f &\longmapsto & U_0f:={\widehat f} \, . 
\end{eqnarray} 
The action of $U_0$ can be written as an integral operator: 
\begin{equation}
      \widehat{f}(E)=U_0f(E)=
       \int_0^{\infty}dr f(r) \overline{\sigma (r;E)} \, , \quad
      f(r)\in L^2([0,\infty ),dr) \, .
      \label{0inteexpre}
\end{equation}
The image of ${\cal D}(H_0)$ under the action of $U_0$ is
\begin{equation}
      {\cal D}(\widehat{H}_0):=U{\cal D}(H_0)=
      \{ {\widehat f}(E) \in L^2((0,\infty ),dE) \, | \ \int_0^{\infty}
                         E^2|\widehat{f}(E)|^2 dE<\infty  \} \, .
\end{equation}

Therefore, we have constructed a unitary operator 
\begin{eqnarray}
      U_0:{\cal D}(H) \subset L^2([0,\infty ),dr) &\longmapsto & 
      {\cal D}(\widehat{H}_0)
       \subset L^2((0,\infty ),dE) \nonumber \\
       f &\longmapsto & \widehat{f}=U_0f
\end{eqnarray}
that transforms from the position representation into the energy 
representation (see diagram (\ref{0diagramsavp}) below). The operator $U_0$ 
diagonalizes the free Hamiltonian in the 
sense that $\widehat{H}_0\equiv U_0H_0U_0^{-1}$ is the multiplication 
operator. The inverse operator of $U_0$ is given by (see Theorem~3 of 
Appendix~\ref{sec:A1})
\begin{equation}
      f(r)=U_0^{-1}\widehat{f}(r)= 
        \int_0^{\infty}dE\, \widehat{f}(E)\sigma (r;E) \, ,
       \quad \widehat{f}(E)\in L^2((0,\infty ),dE) \, .
     \label{0invdiagonaliza}
\end{equation}
The operator $U_0^{-1}$ transforms from the energy representation into the 
position representation (see diagram (\ref{0diagramsavp}) below). The 
expressions (\ref{0inteexpre}) and (\ref{0invdiagonaliza}) provide 
the eigenfunction expansion of any square integrable function in terms of the 
$\delta$-normalized eigensolutions $\sigma (r;E)$ of $h_0$.

\section{Construction of the RHS of the Free Hamiltonian}
\setcounter{equation}{0}
\label{0sec:tneodRHS}

The Sturm-Liouville theory only provides a domain ${\cal D}(H_0)$ on which 
the Hamiltonian $H_0$ is self-adjoint and a unitary operator $U_0$ that 
diagonalizes $H_0$. This unitary operator induces a direct integral 
decomposition of the Hilbert space 
(cf.~Ref.~\cite{DIS} and references therein),
\begin{eqnarray}
      {\cal H} &\longmapsto & 
      U_0{\cal H} \equiv \widehat{\cal H}=
      \oplus \int_{{\rm Sp}(H_0)}{\cal H}(E)dE 
             \nonumber \\
       f &\longmapsto & U_0f\equiv  \{ \widehat{f}(E) \}, \, \quad 
        \widehat{f}(E) \in {\cal H}(E) \, .
       \label{0dirintdec}
\end{eqnarray}

As shown in Refs.~\cite{ANTOINE,DIS,JPA,FP02}, the direct integral 
decomposition does not provide us with all the tools needed in Quantum 
Mechanics. This is why we extend the Hilbert space to the RHS.

We first need to construct a dense invariant domain ${\mathbf \Phi}_0$ on 
which all the powers and all the expectation values of $H_0$ are 
well defined, and on which the Dirac kets act as antilinear functionals. Before
building ${\mathbf \Phi}_0$, we need to build the maximal invariant
subspace ${\cal D}_0$ of $H_0$,
\begin{equation}
      {\cal D}_0:= \bigcap _{n=0}^{\infty}{\cal D}(H_0^n) \, .
      \label{0misus}
\end{equation}
It is easy to check that
\begin{eqnarray}
      {\cal D}_0=\{ \varphi \in L^2([0,\infty ),dr) \, | && \hskip-.5cm  
       h_0^n\varphi (r)\in L^2([0,\infty ),dr),\
       h_0^n\varphi (0)=0,  n=0,1,2,\ldots , \nonumber \\
      && \hskip-.5cm \varphi (r) \in C^{\infty}([0,\infty)) \} \, .
      \label{0mainisexi}
\end{eqnarray}
We can now construct the subspace ${\mathbf \Phi}_0$ on which
the eigenkets $|E\rangle$ of $H_0$ are well defined as antilinear 
functionals. This subspace is given by
\begin{equation}
     {\mathbf \Phi}_0=\{ \varphi \in {\cal D}_0 \, | \ 
     \int_0^{\infty}dr \, \left| (r+1)^n(h_0+1)^m\varphi (r)\right| ^2<\infty, 
     \quad n,m=0,1,2,\ldots \} \, .
\end{equation} 
On ${\mathbf \Phi}_0$, we define the family of norms
\begin{equation}
      \| \varphi \| _{n,m}:= 
    \sqrt{\int_0^{\infty}dr \, \left| (r+1)^n(h_0+1)^m\varphi (r)\right| ^2}
    \ , \quad n,m=0,1,2,\ldots 
      \label{0nmnorms}
\end{equation}
The quantities (\ref{0nmnorms}) fulfill the conditions to be a norm 
(see Proposition~1 of Appendix~\ref{sec:A2}), and can be used to define a 
countably normed topology $\tau _{{\mathbf \Phi}_0}$ on ${\mathbf \Phi}_0$ 
(for the definition of a countably normed topology, consult Ref.~\cite{DIS} 
and references therein),  
\begin{equation}
     \varphi _{\alpha}\, \mapupdown{\tau_{{\mathbf \Phi}_0}}{\alpha \to \infty}
      \, \varphi \quad {\rm iff} \quad  
      \| \varphi _{\alpha}-\varphi \| _{n,m} 
      \, \mapupdown{}{\alpha \to \infty}\, 0 \, , \quad n,m=0,1,2, \ldots 
\end{equation}       
The space ${\mathbf \Phi}_0$ is stable under the action of $H_0$, and 
$H_0$ is $\tau _{{\mathbf \Phi}_0}$-continuous (see Proposition~2 of 
Appendix~\ref{sec:A2}). 

Once we have constructed the space ${\mathbf \Phi}_0$, we can construct its 
topological dual ${\mathbf \Phi}^{\times}_0$ as the space of 
$\tau _{{\mathbf \Phi}_0}$-continuous antilinear functionals on 
${\mathbf \Phi}_0$ and therewith 
the RHS of the free Hamiltonian (see diagram (\ref{0diagramsavp}) below):
\begin{equation}
       {\mathbf \Phi}_0 \subset L^2([0,\infty ),dr) \subset 
       {\mathbf \Phi}^{\times}_0 \, .
\end{equation}

For each $E\in {\rm Sp}(H_0)$, we can now associate a ket $|E\rangle$ to 
the generalized eigenfunction $\sigma (r;E)$ through
\begin{eqnarray}
       |E\rangle :{\mathbf \Phi}_0 & \longmapsto & {\mathbb C} \nonumber \\
       \varphi & \longmapsto & \langle \varphi |E\rangle := 
       \int_0^{\infty}\overline{\varphi (r)}\sigma (r;E) \, dr 
       =\overline{(U_0\varphi )(E)} \, .
       \label{0definitionket}
\end{eqnarray}
The ket $|E\rangle$ in Eq.~(\ref{0definitionket}) is a 
well-defined antilinear functional on ${\mathbf \Phi}_0$, i.e., $|E\rangle$ 
belongs to ${\mathbf \Phi}^{\times}_0$ (see Proposition~3 of 
Appendix~\ref{sec:A2}). The ket 
$|E\rangle$ is a generalized eigenvector of the free Hamiltonian $H_0$ 
(see Proposition~3 of Appendix~\ref{sec:A2}):
\begin{equation}
       H_0^{\times}|E\rangle=E|E\rangle \, ;
\end{equation}
that is,
\begin{equation}
       \langle \varphi |H_0^{\times}|E\rangle=
        \langle H_0^{\dagger}\varphi |E\rangle = E\langle \varphi|E\rangle \, ,
       \quad \forall \varphi \in {\mathbf \Phi}_0 \, .
       \label{0afegenphis}
\end{equation}

\section{The Dirac Basis Vector Expansion for $H_0$}
\setcounter{equation}{0}
\label{sec:0DBVE}

We are now in a position to derive the Dirac basis vector expansion for the
free Hamiltonian. This 
derivation consists of the restriction of the Weyl-Kodaira expansions
(\ref{0inteexpre}) and (\ref{0invdiagonaliza}) to the space 
${\mathbf \Phi}_0$. If we denote 
$\langle r|\varphi \rangle \equiv \varphi (r)$ 
and $\langle E|r\rangle \equiv \overline{\sigma (r;E)}$, and if we define 
the action of the bra $\langle E|$ on $\varphi \in {\mathbf \Phi}_0$ as 
$\langle E| \varphi \rangle := \widehat{\varphi}(E)$, then 
Eq.~(\ref{0inteexpre}) becomes
\begin{equation}
      \langle E|\varphi \rangle =\int_0^{\infty}dr \, 
                                  \langle E |r \rangle
                                  \langle r|\varphi \rangle \, , \quad
       \varphi \in {\mathbf \Phi}_0 \, .
        \label{0DFeps}
\end{equation}
If we denote $\langle r|E \rangle \equiv \sigma (r;E)$, then 
Eq.~(\ref{0invdiagonaliza}) becomes
\begin{equation}
      \langle r|\varphi \rangle =\int_0^{\infty}dE \,
      \langle r|E \rangle  \langle E|\varphi \rangle \, , \quad
       \varphi \in {\mathbf \Phi}_0\, .
       \label{0inveqDva}
\end{equation}
This equation is the Dirac basis vector expansion of the wave function
$\varphi$ in terms of the free eigenkets $|E\rangle$. We can also prove the
Nuclear Spectral Theorem for the free Hamiltonian (see Proposition~4 of 
Appendix~\ref{sec:A2}),
\begin{equation}
      (\varphi ,H_0^n \psi )=\int_0^{\infty}dE \,
      E^n \langle \varphi |E\rangle \langle E|\psi \rangle \, , \quad 
       \forall \varphi ,\psi \in {\mathbf \Phi}_0 \, , n=1,2,\ldots
        \label{0GMT2a}
\end{equation}

\section{Energy Representation of the RHS of $H_0$}
\setcounter{equation}{0}
\label{sec:0ESrepr}

We have already seen that in the energy representation, the Hamiltonian $H_0$
acts as the multiplication operator $\widehat{H}_0$. The energy representation
of the space ${\mathbf \Phi}_0$ is defined as  
\begin{equation}
      \widehat{\mathbf \Phi}_0:= U_0{\mathbf \Phi}_0 \, .
\end{equation}
Obviously $\widehat{\mathbf \Phi}_0$ is a linear subspace
of $L^2([0,\infty ),dE)$. In oder to endow $\widehat{\mathbf \Phi}_0$ with a 
topology $\tau _{\widehat{\mathbf \Phi }_0}$, we carry the topology on 
${\mathbf \Phi}_0$ into $\widehat{\mathbf \Phi}_0$,
\begin{equation}
      \tau _{\widehat{\mathbf \Phi }_0}:=U_0\tau _{{\mathbf \Phi}_0} \, .
\end{equation}
With this topology, the space $\widehat{\mathbf \Phi}_0$ is a linear 
topological space. If we denote the dual space of $\widehat{\mathbf \Phi}_0$
by $\widehat{\mathbf \Phi}_0^{\times}$, then we have
\begin{equation}
      U_0^{\times}{\mathbf \Phi}_0^{\times}=
      (U_0 {\mathbf \Phi}_0)^{\times}= \widehat{\mathbf \Phi}_0^{\times} \, .
\end{equation}
If we denote $|\widehat{E}\rangle \equiv U_0^{\times}|E\rangle$, then we can 
prove that $|\widehat{E}\rangle$ is the antilinear Schwartz delta functional 
(see Proposition~5 of Appendix~\ref{sec:A2}),
\begin{eqnarray}
      |\widehat{E}\rangle: \widehat{\mathbf \Phi} &\longmapsto & 
       {\mathbb C} \nonumber \\
       \widehat{\varphi} &\longmapsto & 
        \langle \widehat{\varphi}|\widehat{E}\rangle :=
        \overline{ \widehat{\varphi}(E)} \, .
\end{eqnarray}

It is very helpful to show the different realizations of the RHS through the
following diagram:
\begin{equation}
      \begin{array}{ccclccclccll}
      H_0; & \varphi (r) & \ & {\mathbf \Phi}_0 & \subset & L^2([0,\infty),dr) &
      \subset & {\mathbf \Phi}_0^{\times} & \  & |E\rangle & \ &
      {\rm position \ repr.} \nonumber \\  [2ex]
       & & \ & \downarrow U_0 &  &\downarrow U_0  &
       & \downarrow U_0^{\times} & \ & & &  \nonumber \\ [2ex]  
      \widehat{H}_0; & \widehat{\varphi}(E) & \ & \widehat{\mathbf \Phi}_0 & 
       \subset & 
      L^2([0,\infty),dE) & \subset & \widehat{\mathbf \Phi}_0 ^{\times} & 
      \ & |\widehat{E}\rangle &\ & {\rm energy \ repr.}  \\ 
      \end{array}
      \label{0diagramsavp}
\end{equation}
On the top line, we have the position representation of the Hamiltonian, the
wave functions, the kets, and the RHS. On the bottom line, we have their
energy representation counterparts.

\section{Conclusions}
\setcounter{equation}{0}
\label{sec:concusios}

We have constructed the RHS of $H_0$ (for the zero angular 
momentum case), and its energy representation. We have associated an eigenket
$|E\rangle$ to each energy $E$ in the spectrum of $H_0$, and shown that
$|E\rangle$ belongs to ${\mathbf \Phi}_0^{\times}$. We have seen
that the energy representation of $|E\rangle$ is given by the antilinear
Schwartz delta functional. We have also shown that the Dirac basis vector
expansion holds within the RHS of $H_0$.

Thus, we conclude that the natural setting for the solutions of the 
Schr\"odinger equation of $H_0$ is the Rigged Hilbert Space 
rather than just the Hilbert space. 

\section*{Acknowledgment}

C.~Koeninger's advise on English style is gratefully acknowledged. This work 
was financially supported by the E.U.~TMR Contract No.~ERBFMRX-CT96-0087 ``The 
Physics of Quantum Information.''

\appendix
\def\thesection{\Alph{section}}
\section{Appendix~A: The Sturm-Liouville Theory}
\setcounter{equation}{0}
\label{sec:A1}

The following theorem provides the procedure to compute the Green 
function of $H_0$ (cf.~Theorem~XIII.3.16 of 
Ref.~\cite{DS} and also Refs.~\cite{DIS,JPA,FP02} for some applications):

\vskip0.5cm

{\bf Theorem~1}\quad  Let $H_0$ be the self-adjoint operator (\ref{0operator}) 
derived from the real formal differential operator (\ref{0doh}) by the 
imposition of the boundary condition (\ref{sac}). Let $\mbox{Im}(E) \neq 0$. 
Then there is exactly one solution $\chi (r;E)$ of $(h_0-E)\sigma =0$ 
square-integrable at $0$ and satisfying the boundary condition (\ref{sac}), 
and exactly one solution $f(r;E)$ of $(h_0-E)\sigma =0$ square-integrable 
at infinity. The resolvent $(E-H_0)^{-1}$ is an integral operator whose kernel 
$G_0(r,s;E)$ is given by
\begin{equation}
       G_0(r,s;E)=\left\{ \begin{array}{ll}
               \frac{2m}{\hbar ^2} \,
      \frac{\chi (r;E) \, f(s;E)}{W(\chi ,f )}
               &r<s \\ [1ex] 
      \frac{2m}{\hbar ^2} \,
      \frac{\chi (s;E) \, f (r;E)}{W(\chi ,f )}
                       &r>s  \, ,
                  \end{array} 
                 \right. 
	\label{exofGFA}
\end{equation}
where $W(\chi ,f )$ is the Wronskian of $\chi$ and $f$
\begin{equation}
       W(\chi ,f )=\chi f'-\chi ' f \, .
\end{equation}

\vskip0.5cm

The theorem that provides the operator $U_0$ that diagonalizes $H_0$ is 
the following (cf.~Theorem XIII.5.13 of Ref.~\cite{DS} and also 
Refs.~\cite{DIS,JPA,FP02} for some applications):

\vskip0.5cm

{\bf Theorem~2} (Weyl-Kodaira)\quad  Let $h_0$ be the formally self-adjoint 
differential operator (\ref{0doh}) defined on the interval 
$[0,\infty )$. Let $H_0$ be the self-adjoint operator (\ref{0operator}). Let
$\Lambda$ be an open interval of the real axis, and suppose that there 
is given a set $\{ \sigma _1(r;E),\, \sigma _2(r;E)\}$ of functions, defined 
and continuous on $(0,\infty )\times \Lambda$, such that for each fixed 
$E$ in $\Lambda$, $\{ \sigma _1(r;E),\, \sigma _2(r;E)\}$ forms a basis for
the space of solutions of $h_0\sigma =E\sigma$. Then there exists a 
positive $2\times 2$ matrix measure $\{ \varrho _{ij} \}$ defined on
$\Lambda$, such that
\begin{enumerate}
      \item the limit 
     \begin{equation}
      (U_0 f)_i(E)=\lim_{c\to 0}\lim_{d\to \infty} 
        \left[ \int_c^d f(r) \overline{\sigma _i(r;E)}dr \right]
      \end{equation}
     exists in the topology of $L^2(\Lambda ,\{ \varrho _{ij}\})$ for each 
    $f$ in $L^2([0,\infty ),dr)$ and defines an isometric isomorphism $U_0$ of
    ${\sf E}(\Lambda )L^2([0,\infty ),dr)$ onto 
       $L^2(\Lambda ,\{ \varrho _{ij}\})$, where ${\sf E}(\Lambda )$ is the
    spectral projection associated with $\Lambda$;
      \item for each Borel function $G$ defined on the real line and vanishing
       outside $\Lambda$,
    \begin{equation}
      U_0{\cal D}(G(H_0))=\{ [f_i]\in L^2(\Lambda ,\{ \varrho _{ij}\}) \, | \
           [Gf_i]\in L^2(\Lambda ,\{ \varrho _{ij}\}) \}
     \end{equation}
    and
    \begin{equation}
      (U_0G(H_0)f)_i(E)=G(E)(U_0f)_i(E), \quad i=1,2, \, E\in \Lambda ,
       \, f\in {\cal D}(G(H_0)) \, .
    \end{equation}  
\end{enumerate}

\vskip0.5cm

The theorem that provides the inverse of the operator $U_0$ is the 
following 
(cf.~Theorem XIII.5.14 of Ref.~\cite{DS} and also Refs.~\cite{DIS,JPA,FP02}
for some applications):

\vskip0.5cm

{\bf Theorem~3} (Weyl-Kodaira)\quad  Let $H_0$, $\Lambda$, 
$\{ \varrho _{ij} \}$, etc., be as in Theorem~2. Let $E_0$ and $E_1$ be the 
end points of $\Lambda$. Then
\begin{enumerate}
      \item the inverse of the isometric isomorphism $U_0$ of 
      ${\sf E}(\Lambda )L^2([0,\infty ),dr)$ onto 
       $L^2(\Lambda ,\{ \varrho _{ij}\})$ is given by the formula
     \begin{equation}
       (U_0^{-1}F)(r)=\lim_{\mu _0 \to E_0}\lim_{\mu _1 \to E_1}
       \int_{\mu _0}^{\mu _1} \left( \sum_{i,j=1}^{2}
             F_i(E)\sigma _j(r;E)\varrho _{ij}(dE) \right)
     \end{equation}
    where $F=[F_1,F_2]\in L^2(\Lambda ,\{ \varrho _{ij}\})$, the limit existing
   in the topology of $L^2([0, \infty ),dr)$;
    \item if $G$ is a bounded Borel function vanishing outside a Borel set 
    $e$ whose closure is compact and contained in $\Lambda$, then $G(H_0)$ has 
    the representation
   \begin{equation}
       G(H_0)f(r)=\int _0^{\infty}f(s)K(H_0,r,s)ds \, , 
   \end{equation}
   where
   \begin{equation}
      K(H_0,r,s)=\sum_{i,j=1}^2 \int_e 
       G(E)\overline{\sigma _i(s;E)}\sigma _j(r;E)\varrho _{ij}(dE) \, .
   \end{equation}
\end{enumerate}

\vskip0.5cm

The spectral measures are provided by the following theorem 
(cf.~Theorem XIII.5.18 of Ref.~\cite{DS} and also Refs.~\cite{DIS,JPA,FP02}
for some applications):

\vskip0.5cm

{\bf Theorem~4} (Titchmarsh-Kodaira)\quad  Let $\Lambda$ be an open interval 
of the real axis and $O$ be an open set in the complex plane containing 
$\Lambda$. Let ${\rm Re}(H_0)$ be the resolvent set of $H_0$. Let 
$\{ \sigma _1(r;E),\, \sigma _2(r;E)\}$ be a set of functions 
which form a basis for the solutions of the equation $h_0\sigma =E\sigma$, 
$E\in O$, and which are continuous on $(0,\infty )\times O$ and analytically 
dependent on $E$ for $E$ in $O$. Suppose that the kernel $G_0(r,s;E)$ for the 
resolvent $(E-H_0)^{-1}$ has a representation
\begin{equation}
      G_0(r,s;E)=\left\{ \begin{array}{lll}
                   \sum_{i,j=1}^2 \theta _{ij}^-(E)\sigma _i(r;E)
                   \overline{\sigma _j(s;\overline{E})}\, , &
                     \qquad & r<s \, ,  \\
                  \sum_{i,j=1}^2 \theta _{ij}^+(E)\sigma _i(r;E)
                   \overline{\sigma _j(s;\overline{E})} \, ,&\qquad & r>s \, ,
                  \end{array}
                 \right.
\end{equation}
for all $E$ in ${\rm Re}(H_0)\cap O$, and that $\{ \varrho _{ij} \}$ is a 
positive matrix measure on $\Lambda$ associated with $H_0$ as in Theorem 2. 
Then the functions $\theta _{ij}^{\pm}$ are analytic in ${\rm Re}(H_0)\cap O$,
and given any bounded open interval $(E_1,E_2)\subset \Lambda$, we have for 
$1\leq i,j\leq 2$,
\begin{equation}
       \begin{array}{lll}
       \varrho _{ij}((E_1,E_2))&=& \lim_{\delta \to 0}\lim_{\varepsilon \to 0+}
         \frac{1}{2\pi i}\int_{E_1+\delta}^{E_2-\delta}
          [ \theta _{ij}^-(E-i\varepsilon )-\theta _{ij}^-(E+i\varepsilon )
          ]dE \\ 
      \quad &=& \lim_{\delta \to 0}\lim_{\varepsilon \to 0+}
         \frac{1}{2\pi i}\int_{E_1+\delta}^{E_2-\delta}
          [ \theta _{ij}^+(E-i\varepsilon )-\theta _{ij}^+(E+i\varepsilon )
          ]dE \, .
         \end{array} 
\end{equation}

\vskip0.5cm

\def\thesection{\Alph{section}}
\section{Appendix~B: Auxiliary Propositions}
\setcounter{equation}{0}
\label{sec:A2}

In this appendix, we list the propositions invoked throughout the paper.

\vskip0.5cm

{\bf Proposition~1} \quad  The quantities
\begin{equation}
    \| \varphi \| _{n,m} := 
    \sqrt{\int_0^{\infty}dr \, \left| (r+1)^n(h_0+1)^m\varphi (r)\right| ^2}
     \ , 
     \quad \varphi \in {\mathbf \Phi}_0 \, , \,  n,m=0,1,2,\ldots, 
      \label{anmnorms}
\end{equation}
are norms.

\vskip0.2cm

{\it Proof}\quad  It is very easy to show that the quantities (\ref{anmnorms}) 
fulfill the conditions to be a norm:
\begin{subequations}
\begin{eqnarray}
      &&\| \varphi +\psi \| _{n,m} \leq \| \varphi \| _{n,m} + 
       \| \psi \| _{n,m} \, , \\
      && \| \alpha \varphi \| _{n,m}=|\alpha |\, \| \varphi \| _{n,m} \, , \\
      && \| \varphi \| _{n,m} \geq 0 \, , \\
      && {\rm If }\  \| \varphi \| _{n,m} =0, \ {\rm then} \ \varphi =0 \, .
        \label{homiensi}
\end{eqnarray}
\end{subequations}
The only condition that is somewhat difficult to prove is (\ref{homiensi}): if 
$\| \varphi \| _{n,m}=0$, then 
\begin{equation}
       (1+r)^n(h_0+1)^m\varphi (r)=0 \, ,
\end{equation}
which yields
\begin{equation}
      (h_0+1)^m\varphi (r)=0 \, . 
      \label{homiodhiis}
\end{equation}
If $m=0$, then Eq.~(\ref{homiodhiis}) implies $\varphi (r)=0$. If $m=1$, then 
Eq.~(\ref{homiodhiis}) implies that $-1$ is an eigenvalue of $H_0$ whose 
corresponding eigenvector is $\varphi$. Since $-1$ is not an
eigenvalue of $H_0$, $\varphi$ must be the zero vector. If $m>1$, the proof 
is similar.

\vskip0.5cm

{\bf Proposition~2}\quad  The space ${\mathbf \Phi}_0$ is stable under the 
action of $H_0$, and $H_0$ is $\tau _{{\mathbf \Phi}_0}$-continuous. 

\vskip0.2cm

{\it Proof}\quad  In order to see that $H_0$ is 
$\tau _{{\mathbf \Phi}_0}$-continuous, we just have to realize that
\begin{eqnarray}
      \| H_0\varphi \| _{n,m}&=&\| (H_0+I)\varphi -\varphi \| _{n,m} 
             \nonumber \\
       &\leq & \| (H_0+I)\varphi \| _{n,m}+ \| \varphi \| _{n,m} \nonumber \\
       &=&\| \varphi \| _{n,m+1}+\| \varphi \| _{n,m} \, .
       \label{tauphiscont}
\end{eqnarray}
We now prove that ${\mathbf \Phi}_0$ is stable under the action of $H_0$. Let
$\varphi \in {\mathbf \Phi}_0$. Saying that $\varphi \in {\mathbf \Phi}_0$ is
equivalent to saying that $\varphi \in {\cal D}_0$ and
that the norms $\| \varphi \| _{n,m}$ are finite for every 
$n,m=0,1,2,\ldots \ $  Since $H_0\varphi$ is also in ${\cal D}_0$, and 
since the 
norms $\| H_0\varphi \| _{n,m}$ are also finite (see Eq.~(\ref{tauphiscont})), 
the vector $H_0\varphi$ is also in ${\mathbf \Phi}_0$.

\vskip0.5cm

{\bf Proposition~3}\quad The function
\begin{eqnarray}
       |E\rangle :{\mathbf \Phi}_0 & \longmapsto & {\mathbb C} \nonumber \\
       \varphi & \longmapsto & \langle \varphi |E\rangle := 
       \int_0^{\infty}\overline{\varphi (r)}\sigma (r;E)dr 
       =\overline{(U_0\varphi )(E)} \, .
       \label{adefinitionket}
\end{eqnarray}
is an antilinear functional on ${\mathbf \Phi}_0$ and a generalized 
eigenvector of (the restriction to ${\mathbf \Phi}_0$ of) $H_0$. 

\vskip0.2cm

{\it Proof}\quad From the definition (\ref{adefinitionket}), it is pretty easy 
to see that $|E\rangle$ is an antilinear functional. In order to show that 
$|E\rangle$ is continuous, we define
\begin{equation}
      {\cal M}(E):= \sup _{r\in [0,\infty )} \left| \sigma (r;E) \right| \, .
\end{equation}
Because
\begin{eqnarray}
      |\langle \varphi |E\rangle | &=& |\overline{U\varphi (E)}| \nonumber \\
      &=&\left| \int_0^{\infty}dr \, \overline{\varphi (r)}\sigma(r;E)\right| 
      \nonumber \\
      &\leq & \int_0^{\infty}dr \, |\overline{\varphi (r)}| |\sigma(r;E)|
      \nonumber \\
      &\leq & {\cal M}(E) \int _0^{\infty}dr \, |\varphi (r)| \nonumber \\
      &=& {\cal M}(E) \int_0^{\infty}dr \,
      \frac{1}{1+r} (1+r) |\varphi (r)| \nonumber \\
      &\leq & {\cal M}(E) \left( \int_0^{\infty}dr \, 
      \frac{1}{(1+r)^2} \right) ^{1/2} 
      \left( \int_0^{\infty}dr \, 
      \left| (1+r) \varphi (r) \right| ^2 \right) ^{1/2} \nonumber \\
      &=&{\cal M}(E) \left( \int_0^{\infty}dr \, 
      \frac{1}{(1+r)^2} \right) ^{1/2} 
       \| \varphi \| _{1,0}     \nonumber \\
       &=&{\cal M}(E) \| \varphi \| _{1,0} \, ,
\end{eqnarray}
the functional $|E\rangle$ is continuous when ${\mathbf \Phi}_0$ is endowed 
with the $\tau _{{\mathbf \Phi}_0}$ topology.

In order to prove that $|E\rangle$ is a generalized eigenvector of $H_0$, we 
make use of the conditions (\ref{0mainisexi}) and (\ref{0nmnorms})
satisfied by the elements of ${\mathbf \Phi}_0$:
\begin{eqnarray}
       \langle \varphi |H_0^{\times}|E\rangle &=& 
          \langle H_0^{\dagger}\varphi |E\rangle
       \nonumber \\
       &=& \int_0^{\infty}dr \, 
       \left( -\frac{\hbar ^2}{2m}\frac{d^2}{dr^2} 
       \overline{\varphi(r)} \right) \sigma (r;E) \nonumber \\
       &=&-\frac{\hbar ^2}{2m}
       \left[ \frac{d\overline{\varphi (r)}}{dr} \sigma(r;E) 
       \right] _0^{\infty} 
       +\frac{\hbar ^2}{2m}
       \left[ \overline{\varphi (r)} \frac{d\sigma(r;E)}{dr} 
       \right] _0^{\infty} \nonumber \\ 
       &&+ \int_0^{\infty}dr \, \overline{\varphi(r)}
       \left( -\frac{\hbar ^2}{2m}\frac{d^2}{dr^2} \sigma (r;E) \right)
        \nonumber \\
       &=&E\langle \varphi |E\rangle \, .
\end{eqnarray}

Similarly, one can also prove that
\begin{equation}
       \langle \varphi | (H_0^{\times})^n|E\rangle =
       E^n \langle \varphi |E\rangle \, .
\end{equation}

\vskip0.5cm

{\bf Proposition~4} (Nuclear Spectral Theorem) \quad Let 
\begin{equation}
      {\mathbf \Phi}_0 \subset L^2([0,\infty ),dr)\subset 
      {\mathbf \Phi}_0^{\times}
\end{equation}
be the RHS of $H_0$ such that ${\mathbf \Phi}_0$ 
remains invariant under $H_0$ and $H_0$ is a 
$\tau _{{\mathbf \Phi}_0}$-continuous 
operator on ${\mathbf \Phi}_0$. Then, for each $E$ in 
the spectrum of $H_0$ there is a generalized eigenvector $|E\rangle$ such that
\begin{equation}
       H_0^{\times}|E\rangle =E|E\rangle
\end{equation}
and such that
\begin{equation}
      (\varphi ,\psi )=\int_{ {\rm Sp}(H_0)}dE\, 
      \langle \varphi |E\rangle \langle E|\psi \rangle \, , \quad 
       \forall \varphi ,\psi \in {\mathbf \Phi}_0 \, ,
       \label{GMT1}
\end{equation}
and
\begin{equation}
      (\varphi ,H_0^n \psi )=\int_{ {\rm Sp}(H_0)}dE \,
      E^n \langle \varphi |E\rangle \langle E|\psi \rangle \, , \quad 
       \forall \varphi ,\psi \in {\mathbf \Phi}_0 \, , n=1,2,\ldots
        \label{GMT2}
\end{equation}

\vskip0.2cm

{\it Proof} \quad Let $\varphi$ and $\psi$ be in ${\mathbf \Phi}_0$. Since 
$U_0$ is unitary,
\begin{equation}
       (\varphi ,\psi )=(U_0\varphi ,U_0\psi )=
         (\widehat{\varphi} ,\widehat{\psi})  \, .
       \label{Usiuni}
\end{equation}
The wave functions $\widehat{\varphi}$ and $\widehat{\psi}$ are in particular
elements of $L^2([0,\infty ),dE)$. Therefore their scalar product is 
well defined,
\begin{equation}
      (\widehat{\varphi} ,\widehat{\psi} )=
      \int_{{\rm Sp}(H_0)}dE \, 
      \overline{ \widehat{\varphi}(E)} \widehat{\psi}(E) \, .
      \label{sphatvhaps}
\end{equation}
Because $\varphi$ and $\psi$ belong to ${\mathbf \Phi}_0$, the action of each 
eigenket $|E\rangle$ on them is well defined,
\begin{subequations}
         \label{actionofEpsi}
\begin{eqnarray}
      \langle \varphi |E\rangle =\overline{ \widehat{\varphi}(E)} \, ,\\
      \langle E|\psi \rangle =\widehat{\psi}(E) \, .
\end{eqnarray}
\end{subequations}
By plugging Eq.~(\ref{actionofEpsi}) into Eq.~(\ref{sphatvhaps}) and
Eq.~(\ref{sphatvhaps}) into Eq.~(\ref{Usiuni}), we obtain 
Eq.~(\ref{GMT1}). The proof of (\ref{GMT2}) is similar:
\begin{eqnarray}
       (\varphi ,H_0^n\psi )&=&(U_0\varphi , U_0H_0^nU_0^{-1}U_0\psi ) 
              \nonumber \\
       &=& (\widehat{\varphi} ,\widehat{H}_0^n\widehat{\psi} ) \nonumber \\
       &=&\int_{{\rm Sp}(H_0)}dE \,\overline{ \widehat{\varphi}(E)} 
        (\widehat{H}_0^n\widehat{\psi})(E) \nonumber \\
       &=&\int_{{\rm Sp}(H_0)}dE\, E^n \overline{ \widehat{\varphi}(E)} 
        \widehat{\psi}(E) \nonumber \\
       &=& \int_{{\rm Sp}(H_0)}dE \,
      E^n \langle \varphi |E\rangle \langle E|\psi \rangle  \, . 
\end{eqnarray}

\vskip0.5cm

{\bf Proposition~5} \quad The energy representation of the eigenket 
$|E\rangle$ is the antilinear Schwartz delta functional $|\widehat{E}\rangle$.

\vskip0.2cm

{\it Proof} \quad Because 
\begin{eqnarray}
       \langle \widehat{\varphi }|U_0^{\times}|E\rangle &=&
       \langle U_0^{-1}\widehat{\varphi }|E\rangle  \nonumber \\
       &=& \langle \varphi |E\rangle \nonumber \\
       &=& \int_0^{\infty}\overline{\varphi (r)}\sigma (r;E)dr \nonumber \\
       &=& \overline{ \widehat{\varphi}(E)} \, ,
\end{eqnarray}
the functional $U_0^{\times}|E\rangle =|\widehat{E}\rangle$ is the antilinear
Schwartz delta functional.


\begin{thebibliography}{8.}
\bibitem{SL} We shall use the Sturm-Liouville theory in the form that 
appears in the treatise of Dunford and Schwartz~\cite{DS}. Applications of 
the Sturm-Liouville theory to simple potentials can be found in 
Refs.~\cite{JPA,FP02,BRANDAS1,BRANDAS,CSF}.
\end{thebibliography}
\end{document}